\documentclass[aps, preprint, groupedaddress, floatfix]{revtex4}

\usepackage{epsfig}
\usepackage{graphicx}
\usepackage{amsmath}

\begin{document}

\title{Charge transfer across transition metal oxide interfaces: emergent
conductance and new electronic structure}

\author{Hanghui Chen$^{1,2}$, Hyowon Park$^{1,2}$, Andrew J. Millis$^{1}$ and Chris A. Marianetti$^{2}$ }

\affiliation{
 $^1$Department of Physics, Columbia University, New York, NY, 10027, USA\\
 $^2$Department of Applied Physics and Applied Mathematics, Columbia University, New York, NY, 10027,
 USA\\}
\date{\today}

\begin{abstract}
We perform density functional theory plus dynamical mean field theory
calculations to investigate internal charge transfer in an artificial
superlattice composed of alternating layers of vanadate and manganite
perovskite and Ruddlesden-Popper structure materials. We show that the
electronegativity difference between vanadium and manganese
causes moderate charge transfer from VO$_2$ to MnO$_2$ layers in
both perovskite and Ruddlesden-Popper based superlattices, leading to
hole doping of the VO$_2$ layer and electron doping of the MnO$_2$
layer. Comparison of the perovskite and Ruddlesden-Popper based
heterostructures provides insights into the role of the apical oxygen.
Our first principles simulations demonstrate that the combination of
internal charge transfer and quantum confinement provided by
heterostructuring is a powerful approach to engineering electronic
structure and tailoring correlation effects in transition metal
oxides.
\end{abstract}

\maketitle

\section{Introduction}

Advances in thin film epitaxy growth techniques have made it possible
to induce emergent electronic~\cite{Hwang-Nature-2002,
  Okamoto-Nature-2004, Moetakef-APL-2011,
  Hwang-Nature-2004,Mannhart-Science-2006, Mannhart-Science-2007,
  Triscone-Nature-2008, Koida-PRB-2002,
  Smadici-PRL-2007,Nanda-PRL-2008},
magnetic~\cite{Brinkman-NatMat-2007,Bert-NatPhys-2011,
  Li-NatPhys-2011} and
orbital~\cite{Chakhalian-Science-2007,Benckiser-NatMater-2011} states,
which are not naturally occurring in bulk constituents, at atomically
sharp transition metal oxide interfaces~\cite{Hwang-Science-2006,
  Tokura-NatMat-2008, Mannhart-Science-2010,
  Chakhalian-NatMat-2012}. For example, the interface between the two
nonmagnetic band insulators LaAlO$_3$ and
SrTiO$_3$~\cite{Hwang-Nature-2004} has been reported to exhibit both
conductance~\cite{Hwang-NatMat-2006} and
magnetism~\cite{Lee-NatMat-2013} (see reviews~\cite{Mannhart-MRS-2008,
  Huijben-AM-2009, Chen-Adv-2010,Zubko-ARCMP-2011} and references
therein). At the interface of Mott insulators SrMnO$_3$ and
LaMnO$_3$, hole doping on the Mn sites leads to rich phenomena, including
metal-insulator transition, charge/spin/orbital ordering and
magnetoresistance~\cite{Salamon-RevModPhys-2001, Moreo-PRL-2000, Bhattacharya-PRL-2008}.

In LaAlO$_3$/SrTiO$_3$ and related heterostructures, the interface
electron gas is believed to be produced by the polar catastrophe
mechanism, which leads to the transfer of charge from the sample
surface to the interface. Here, we consider a different mechanism
for controlling the electronic properties of an interface: namely,
electronegativity-driven charge transfer. Recently, we have shown
that internal charge transfer in a LaTiO$_3$/LaNiO$_3$
superlattice transforms metallic LaNiO$_3$ into a $S=1$ Mott
insulator and Mott insulating LaTiO$_3$ into a $S=0$ band
insulator~\cite{Chen-PRL-2013b}.  A natural question arises: can
we reverse the process and utilize internal charge transfer to
induce conductance via oxide interfaces? In this regard, it is
very tempting to explore Mott interfaces (one or both constituents
are Mott insulators) due to the unusual phenomena (colossal
magnetoresistance and high temperature superconductivity)
exhibited in certain doped Mott insulators.

In this paper we use density functional theory + dynamical mean field
theory (DFT+DMFT) to theoretically design a superlattice with emergent
metallic behavior. We explore two different types of structure: the
perovskite structure (referred to as 113-type) and the $n=1$
Ruddlesden-Popper structure (referred to as 214-type). Among the four
bulk constituents (SrVO$_3$, SrMnO$_3$, Sr$_2$VO$_4$ and
Sr$_2$MnO$_4$), all are correlation-driven insulators except SrVO$_3$,
which is a moderately correlated metal~\cite{Matsuno-PRL-2005,
  Sondenaa-PRB-2006, Dang-PRB-2014}. We show that the difference of
electronegativity between the elements V and Mn drives internal charge
transfer from V to Mn sites, leading to a ``self-doping'' at the
interface and possibly inducing conductance as Mn sites become weakly
electron doped and V sites hole doped.

The rest of the paper is organized as follows: Section II presents
the theoretical methods. A schematic of band alignment is
presented in Section III to illustrate the underlying mechanism of
charge transfer. All the bulk results from \textit{ab initio}
calculations are in Section IV and the results of
vanadate-manganite superlattices are in Section V, both of which
provide qualitative support and quantitative corrections to the
schematic.  The conclusions are in Section VI. Five Appendices
present technical details relating to the insulating gaps of
Sr$_2$VO$_4$ and Sr$_2$MnO$_4$, alternative forms of the double
counting correction, LDA spectra of the superlattices and the
possibility of two consecutive repeating layers (i.e. 2/2
superlattices instead of 1/1 superlattices).

\section{Computational details}

\begin{figure}[t]
\includegraphics[angle=-90,width=0.9\textwidth]{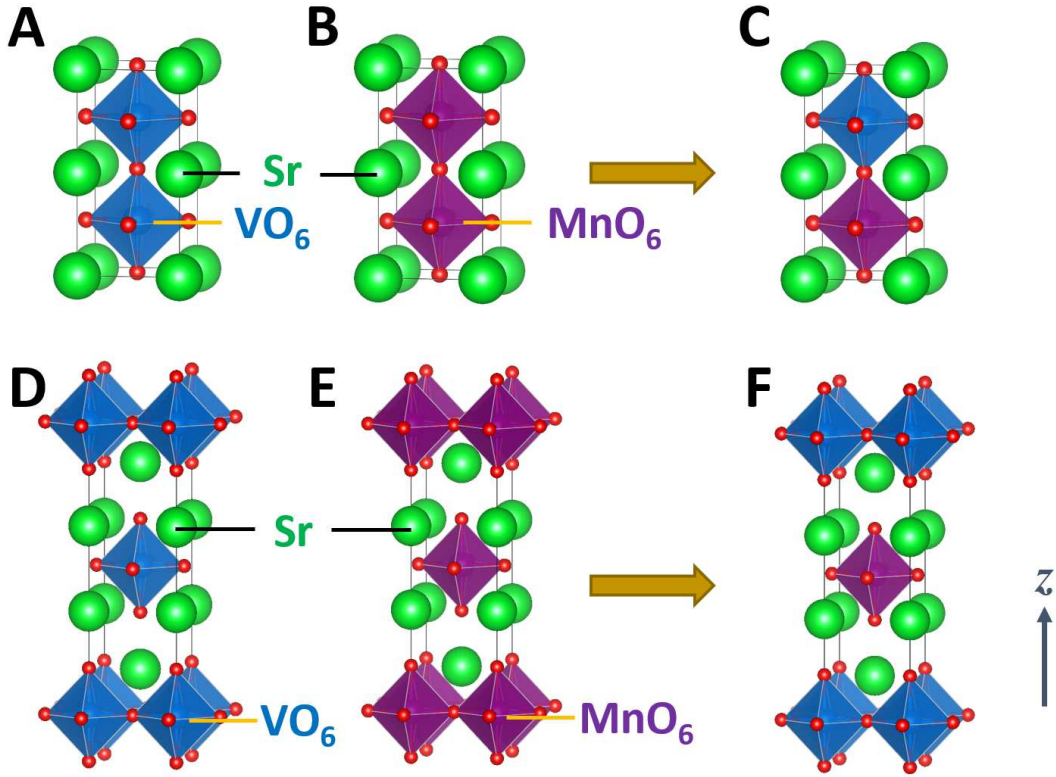}
\caption{\label{fig:structure} Simulation cells of \textbf{A})
bulk SrVO$_3$, \textbf{B}) bulk SrMnO$_3$ and \textbf{C})
SrVO$_3$/SrMnO$_3$ superlattice; \textbf{D}) bulk Sr$_2$VO$_4$,
\textbf{E}) bulk Sr$_2$MnO$_4$ and \textbf{F})
Sr$_2$VO$_4$/Sr$_2$MnO$_4$ superlattice. The green atoms are Sr.
The blue and purple cages are VO$_6$ and MnO$_6$ octahedra,
respectively. The stacking direction of the superlattice is the
[001] axis.}
\end{figure}

The DFT~\cite{Hohenberg-PR-1964, Kohn-PR-1965} component of our
DFT+DMFT~\cite{Georges-RMP-1996, Kotliar-RMP-2006} calculations is
performed using a plane-wave basis~\cite{Payne-RMP-1992}, as
implemented in the Vienna Ab-initio Simulation Package
(VASP)~\cite{Kresse1993558,Kresse19948245,Kresse199615,Kresse-PRB-1996}
using the Projector Augmented Wave (PAW)
approach~\cite{Blochl-PRB-1994,Kresse-PRB-1999}. Both local density
approximation (LDA)~\cite{Perdew-PRB-1981} and Perdew-Burke-Ernzerhof
generalied gradient approximation (GGA-PBE)~\cite{Perdew-PRL-1996} are
employed.  The correlated subspace and the orbitals with which it
mixes are constructed using maximally localized Wannier
functions~\cite{Marzari-RMP-2012} defined over the full 10 eV range
spanned by the $p$-$d$ band complex, resulting in an well localized
set of $d$-like orbitals~\cite{wann}. To find the stationary solution
for our DFT+DMFT functional, we first find the self-consistent charge
density within DFT. Subsequently we fix the charge density and
converge the DMFT equations. A full charge self-consistency is not
implemented in the present work. However, this approximation procedure
is found to yield reasonable results in calculations of bulk
systems~\cite{Wang-PRB-2012, Dang-PRB-2014, Park-PRB-2014}.

For the bulk materials, we consider two structures: the experimental
one and the theoretical relaxed structure obtained by the use of
DFT. For the superlattice, we use DFT to obtain the relaxed
structure. We compare LDA and GGA calculations and both exchange
correlation functions yield consistent results.  The simulation cell
is illustrated in Fig.~\ref{fig:structure}. The stacking direction of
the superlattice is along [001]. We use an energy cutoff 600 eV. A
$12\times 12\times 12 \left[\frac{L_x}{L_z}\right]$ ($L_x$ and $L_z$
are the lattice constants along the $x$ and $z$ directions, and [$x$]
is the integer part of $x$) Monkhorst-Pack grid is used to sample the
Brillouin zone. Both cell and internal coordinates are fully relaxed
until each force component is smaller than 10 meV/\AA~and the stress
tensor is smaller than 10 kBar. Convergence of the key results are
tested with a higher energy cutoff (800 eV) and a denser $k$-point
sampling $20\times 20\times 20\left[\frac{L_x}{L_z}\right]$ and no
significant changes are found.

For the vanadates, we treat the empty $e_g$ orbitals with a static
Hartree-Fock approximation (recent work shows this approximation is
adequate to describe the electronic structure of vanadates
~\cite{Dang-PRB-2014}), while correlations in the V $t_{2g}$ manifold
are treated within single-site DMFT including the Slater-Kanamori
interactions using intra-orbital Hubbard $U_{\textrm{V}}$ = 5 eV and
$J_{\textrm{V}}$ = 0.65 eV~\cite{Pavarini-PRL-2004, Nekrasov-PRB-2005,
  Aichhorn-PRB-2009}. For manganites, we treat the correlations on all
the five Mn $d$ orbitals within single-site DMFT using the
Slater-Kanamori interactions with intra-orbital Hubbard
$U_{\textrm{Mn}}$ = 5 eV and $J_{\textrm{Mn}}$ = 1
eV~\cite{Park-PRL-1996}.  The DMFT impurity problem is solved using
the continuous time quantum Monte Carlo method ~\cite{Werner-PRL-2006,
  Werner-PRB-2006, Haule-PRB-2007}. In order to use the ``segment''
algorithm~\cite{Gull-RMP-2011}, we neglect the exchange and pairing
terms in the Slater-Kanamori Hamiltonian.  All the calculations are
paramagnetic and the temperature is set to 232 K.  Long-range magnetic
ordering (in particular antiferromagnetism) might be induced at low
temperature on Mn sites in manganites and in superlattices. A thorough
study of magnetic properties will be presented
elsewhere~\cite{Chen-unpublished}. For the superlattice, we solve the
problem in the single-site DMFT approximation, meaning that the self
energy is site local and is one function on the V site and a different
one on the Mn site. The self energies are determined from two quantum
impurity models, which are solved independently but coupled at the
level of the self consistency condition.

An important outstanding issue in the DFT+DMFT procedure is the
``double counting correction'' which accounts for the part of the
Slater-Kanamori interactions already included in the underlying
DFT calculation and plays an important role by setting the mean
energy difference between the $d$ and $p$ bands. The $p$-$d$
separation plays a crucial role in determining the band alignment,
which affects the charge transfer.  However, currently there is no
exact procedure for the double counting correction.  We use the
$U'$ double counting method recently
introduced~\cite{Park-PRB-2014}, where the parameter $U'$ is the
prefactor in the double-counting which determines the $p$-$d$
separation and equivalently the number of electrons in the
$d$-manifold.  In this study, $U'$ is chosen to produce an energy
separation between the O $p$ and transition metal $d$ bands which
is consistent with photoemission experiments. Our main qualitative
conclusions do not depend on the details of the double counting
scheme; in particular we show here they hold also for the
conventional fully localized limit (FLL) double
counting~\cite{Czyzyk-PRB-1994} which is the $U'=U$ limit of the
method of Ref.~\cite{Park-PRB-2014}. The reason behind that is
because in the superlattice, it is the relative V$d$-Mn$d$ energy
separation that controls the charge transfer. The FLL double
counting formula underestimates the $p$-$d$ separation in both
SrVO$_3$ and SrMnO$_3$ by about 1 eV.  However, such an error is
cancelled in the calculation of V$d$-Mn$d$ energy separation.
Therefore the FLL double counting does not change the charge
transfer picture.

The spectral function presented throughout this work is defined as follows:

\begin{equation}
\label{eqn:spectral} A_i(\omega) = -\frac{1}{\pi N_k}\sum_{\textbf{k}}
\textrm{Im}\left(\left[(\omega+\mu)\textbf{I}-H_0(\textbf{k})-\Sigma_{\textrm{tot}}(\omega)+ V_{dc})\right]^{-1}\right)_{ii}
\end{equation}
where $i$ is the label of a Wannier function, $N_k$ is the number of
$k$-points, $\textbf{I}$ is an identity matrix, $H_0(\textbf{k})$ is
the DFT-LDA band Hamiltonian in the matrix form using the Wannier
basis. $\Sigma_{\textrm{tot}}(\omega)$ is the total self-energy and is
understood as a diagonal matrix only with nonzero entries on the
correlated orbitals. Local tetragonal point symmetry of the V and Mn
sites ensures that $\Sigma(\omega)$ is diagonal within the correlated
orbital subspace. $\mu$ is the chemical potential. $V_{dc}$ is the
double counting potential, which is defined as~\cite{Park-PRB-2014}:

\begin{equation}
\label{eqn:dc} V_{dc} = (U' - 2J)\left(N_d - \frac{1}{2}\right) - \frac{1}{2}J(N_d - 3)
\end{equation}
Note that if $U'=U$, then we restore the standard FLL double
counting formula~\cite{double}. For clarity, all the spectra
functions presented in this paper are obtained from LDA+DMFT
calculations. GGA+DMFT calculations yield qualitatively consistent
results.

\section{Schematic of band structure and band alignment}

\begin{figure}[t]
\includegraphics[angle=-90,width=0.9\textwidth]{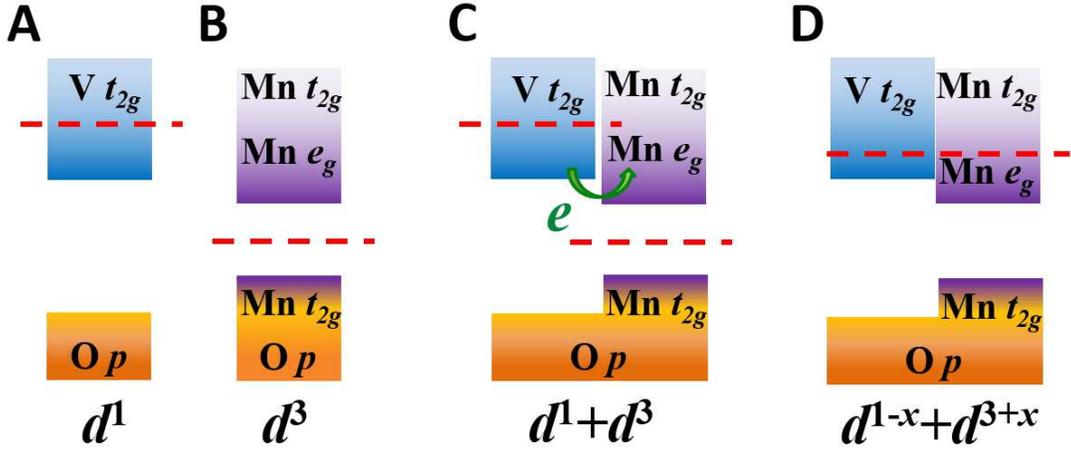}
\caption{\label{fig:band} Schematic band structure of
\textbf{A}) vanadates and \textbf{B}) manganites. \textbf{C}) is
the band alignment of the superlattice before the charge transfer occurs,
i.e. two independent Fermi levels. \textbf{D}) is the band structure
of the superlattice after the charge transfer occurs, i.e. with one common
Fermi level. The dashed red line denotes the Fermi level.}
\end{figure}

We consider the following materials as components of the superlattice:
SrVO$_3$, a moderately correlated metal with nominal $d$-valence
$d^1$; Sr$_2$VO$_4$, a correlation-driven insulator also with nominal
valence $d^1$; and SrMnO$_3$ and Sr$_2$MnO$_4$, both of which are
$d^3$ correlation-driven (Mott) insulators. Fig.~\ref{fig:structure}
shows the atomic structure of the bulk phases of the constituent
materials and the corresponding
superlattices. Fig.~\ref{fig:structure}\textbf{A}, \textbf{B}, and
\textbf{C} are bulk SrVO$_3$, bulk SrMnO$_3$, and SrVO$_3$/SrMnO$_3$
superlattice, respectively. Fig.~\ref{fig:structure}\textbf{D},
\textbf{E}, and \textbf{F} are bulk Sr$_2$VO$_4$, bulk Sr$_2$MnO$_4$,
and Sr$_2$VO$_4$/Sr$_2$MnO$_4$ superlattice, respectively. In both
superlattices, the stacking direction is along the [001] axis. In the
214-type, the V atoms are shifted by a
$\left(\frac{1}{2},\frac{1}{2}\right)$ lattice constant in the $xy$
plane relative to the Mn atoms.

Fig.~\ref{fig:band} is a schematic of the band structure of bulk
vanadates, bulk manganites, and the band alignments in the
superlattice (the small insulating gap of Sr$_2$VO$_4$ is not relevant
here.  There is a large energy separation (around 2 eV) between V $d$
and O $p$ states (see Fig.~\ref{fig:band}\textbf{A}).  In the Mn-based
materials (see Fig.~\ref{fig:band}\textbf{B}), the highest occupied
states are Mn $t_{2g}$-derived and the lowest unoccupied states are Mn
$e_g$-derived. Due to the electronegativity difference between V and
Mn, visible as the difference in the energy separation of the
transition metal $d$ levels from the oxygen $p$ levels, if we align
the O $p$ states between vanadates and manganites (see
Fig.~\ref{fig:band}\textbf{C}), the occupied V $t_{2g}$ states overlap
in energy with the unoccupied Mn $e_g$ states. The overlap drives
electrons from V sites to Mn sites. As the superlattice is formed, a
common Fermi level appears across the interface and thus we expect
that Mn $e_g$ states become electron doped and V $t_{2g}$ states hole
doped.

We make two additional points: i) though SrVO$_3$ is a metal and
Sr$_2$VO$_4$ is an insulator with a small energy gap (around 0.2
eV)~\cite{Matsuno-PRL-2005}, the near Fermi level electronic structure
does not affect the band alignment and therefore the internal charge
transfer is expected to occur no matter whether there is a small
energy gap in V $t_{2g}$ states at the Fermi level or not; ii) in our
schematic, we assume that the main peak of O $p$ states are exactly
aligned between the vanadates and manganites in the superlattices.  Of
course, real material effects will spoil any exact alignment.  We will
use \textit{ab initio} calculations to provide quantitative
information on how O $p$ states are aligned between the two materials.

\section{Bulk properties}

\begin{table}[t]
\caption{\label{tab:structure} The in-plane and out-of-plane V-O and
  Mn-O bond lengths $l$ of SrVO$_3$, SrMnO$_3$, Sr$_2$VO$_4$ and
  Sr$_2$MnO$_4$. The corresponding VO$_6$ and MnO$_6$ octahedral
  volumes $\Omega$ are also calculated. The relaxed structures are
  obtained from DFT-LDA and DFT-GGA non-spin-polarized
  calculations. The experimental values which are referenced in the main text
  are also provided for comparison.}
\begin{center}
\begin{tabular}{c|c|c|c|c|c|c|c|c}
\hline
\hline
       &  \multicolumn{3}{c|}{SrVO$_3$} & \multicolumn{3}{c|}{SrMnO$_3$} & \multicolumn{2}{c}{SrVO$_3$/SrMnO$_3$} \\
\hline &  LDA & GGA & exp & LDA & GGA & exp & LDA & GGA \\
\hline
$l_{\textrm{in}}$(V-O) & 1.89~\AA~& 1.93~\AA & 1.92~\AA & \multicolumn{3}{c|}{--} & 1.88~\AA & 1.92~\AA \\
\hline
$l_{\textrm{out}}$(V-O) & 1.89~\AA~& 1.93~\AA & 1.92~\AA & \multicolumn{3}{c|}{--} & 1.85~\AA & 1.88~\AA \\
\hline
$\Omega_{\textrm{VO}_6}$ & 9.00~\AA$^3$ & 9.59~\AA$^3$ & 9.44~\AA$^3$ & \multicolumn{3}{c|}{--} & 8.72~\AA$^3$ & 9.24~\AA$^3$ \\
\hline
$l_{\textrm{in}}$(Mn-O) & \multicolumn{3}{c|}{--} & 1.86~\AA~ & 1.90~\AA & 1.90~\AA & 1.88~\AA & 1.92~\AA  \\
\hline
$l_{\textrm{out}}$(Mn-O) & \multicolumn{3}{c|}{--} & 1.86~\AA~ & 1.90~\AA & 1.90~\AA & 1.89~\AA & 1.94~\AA \\
\hline
$\Omega_{\textrm{MnO}_6}$ & \multicolumn{3}{c|}{--} & 8.58~\AA$^3$ & 9.15~\AA$^3$ & 9.15~\AA$^3$ & 8.91~\AA$^3$ & 9.54~\AA$^3$ \\
\hline\hline
&  \multicolumn{3}{c|}{Sr$_2$VO$_4$} & \multicolumn{3}{c|}{Sr$_2$MnO$_4$} & \multicolumn{2}{c}{Sr$_2$VO$_4$/Sr$_2$MnO$_4$} \\
\hline
 &  LDA & GGA & exp & LDA & GGA & exp & LDA & GGA \\
\hline
$l_{\textrm{in}}$(V-O)  & 1.88~\AA~ & 1.92~\AA & 1.91~\AA & \multicolumn{3}{c|}{--} & 1.85~\AA & 1.90~\AA \\
\hline
$l_{\textrm{out}}$(V-O) & 1.96~\AA~ & 2.00~\AA & 1.95~\AA & \multicolumn{3}{c|}{--} & 1.93~\AA & 1.95~\AA \\
\hline
$\Omega_{\textrm{VO}_6}$ & 9.24~\AA$^3$ & 9.83~\AA$^3$ & 9.49~\AA$^3$ & \multicolumn{3}{c|}{--} & 8.81~\AA$^3$ & 9.39~\AA$^3$ \\
\hline
$l_{\textrm{in}}$(Mn-O)  & \multicolumn{3}{c|}{--}  & 1.82~\AA & 1.86~\AA & 1.90~\AA & 1.85~\AA & 1.90~\AA \\
\hline
$l_{\textrm{out}}$(Mn-O) & \multicolumn{3}{c|}{--} & 1.99~\AA~ & 2.04~\AA & 1.95~\AA & 1.99~\AA & 2.06~\AA \\
\hline
$\Omega_{\textrm{MnO}_6}$ & \multicolumn{3}{c|}{--} & 8.79~\AA$^3$ & 9.41~\AA$^3$ & 9.39~\AA$^3$ & 9.08~\AA$^3$ & 9.92~\AA$^3$ \\
\hline\hline
\end{tabular}
\end{center}
\end{table}

This section is devoted to properties of vanadates and manganites in
their bulk single crystalline form. We perform DFT+DMFT calculations
on both experimental structures and relaxed atomic structures
obtained from DFT-LDA. The DFT-LDA relaxed V-O and Mn-O bond
lengths, as well as the volume of VO$_6$ and MnO$_6$ octahedra, are
summarized in Table~\ref{tab:structure}, along with the experimental
bond lengths and octahedral volumes (in parentheses) for
comparison. However, in order to directly compare to the photoemission
data, we only present the spectral functions that are calculated using
the experimental structures.

\subsection{Bulk vanadates}

\begin{figure}[t]
\includegraphics[angle=-90,width=0.9\textwidth]{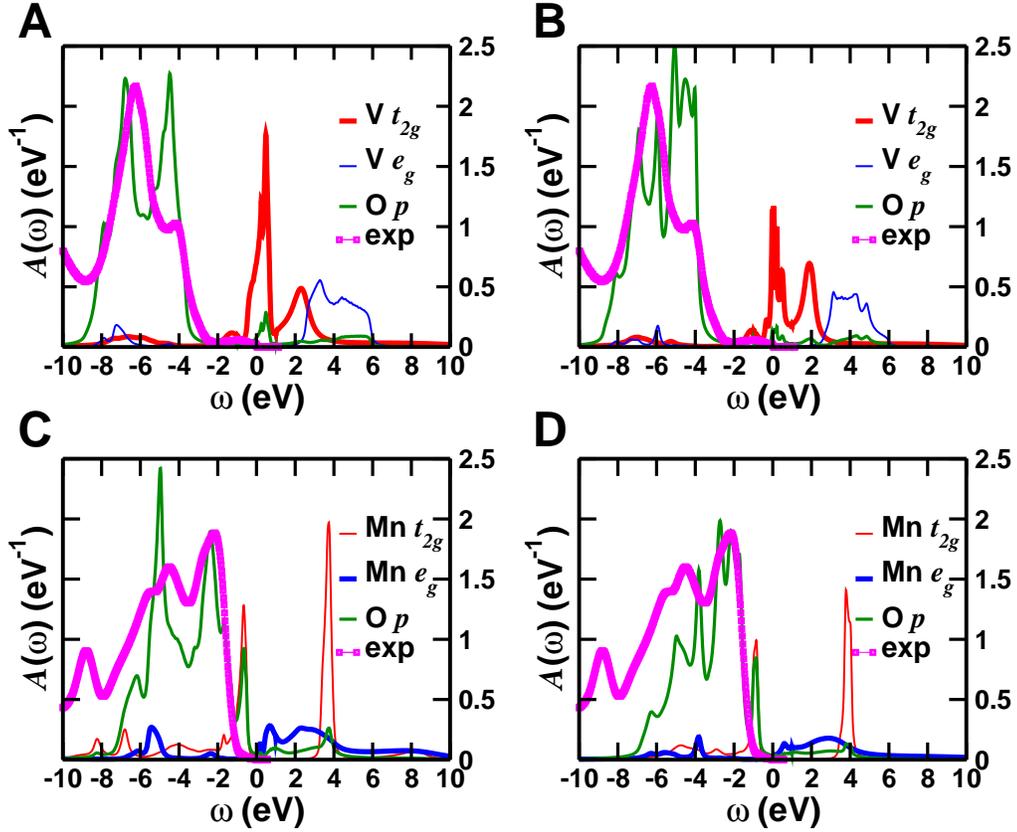}
\caption{\label{fig:bulk} Orbitally resolved spectral function of
  \textbf{A}) SrVO$_3$ and \textbf{B}) Sr$_2$VO$_4$; \textbf{C})
  SrMnO$_3$ and \textbf{D}) Sr$_2$MnO$_4$, obtained from LDA+DMFT
  calculations. The pink dots are the experimental spectra for either
  SrVO$_3$ or SrMnO$_3$ (identical data are plotted alongside the
  theoretical spectra for the Ruddlesden-Popper
  structures)~\cite{cornell}. For vanadates, $U'$ double counting is
  employed with $U_{\textrm{V}}$ = 5 eV and $U'_{\textrm{V}}$ = 3.5
  eV. The red (very thick), blue (thin) and green (thick) curves are V
  $t_{2g}$, V $e_g$ and O $p$ projected spectral functions,
  respectively.  For manganites, $U'$ double counting is employed with
  $U_{\textrm{Mn}}$ = 5 eV and $U'_{\textrm{Mn}}$ = 4.5 eV. The red
  (thin), blue (very thick) and green (thick) curves are Mn $t_{2g}$,
  Mn $e_g$ and O $p$ projected spectral functions, respectively.  The
  Fermi level is set at zero energy.}
\end{figure}

We begin with bulk vanadates: SrVO$_3$ and Sr$_2$VO$_4$. SrVO$_3$ has
a cubic structure with a lattice constant $a$ = 3.841~\AA
~\cite{Maekawa-JAC-2006}. Sr$_2$VO$_4$ forms $n=1$ Ruddlesden-Popper
structure with the in-plane lattice constant $a$ = 3.826~\AA~and the
out-of-plane lattice constant $c$ =
12.531~\AA~\cite{Zhou-PRL-2007}. We use a Hubbard
$U_{\textrm{V}}$ = 5 eV on both vanadate materials to include correlation
effects on V $d$ orbitals, which is in the vicinity of previous
studies~\cite{Pavarini-PRL-2004, Nekrasov-PRB-2005, Aichhorn-PRB-2009}.

Fig.~\ref{fig:bulk} shows the orbitally-resolved spectral function
$A(\omega)$ of bulk SrVO$_3$ (Fig.~\ref{fig:bulk}\textbf{A}) and bulk
Sr$_2$VO$_4$ (Fig.~\ref{fig:bulk}\textbf{B}), along with the
experimental photoemission data for bulk SrVO$_3$~\cite{cornell}.
The threshold of O $p$ states is around 2 eV below
the Fermi level.  We find that $U'_{\textrm{V}}$ = 3.5 eV yields a
reasonable agreement between the calculated O $p$ states and
experimental photoemission data. At $U_{\textrm{V}}$ = 5 eV,
with the $p$-$d$ separation fixed by the experimental photoemission
data, our DFT+DMFT calculations find SrVO$_3$ to be metallic, consistent
with the experiment. However, they do not reproduce a Mott insulating
state in Sr$_2$VO$_4$, as observed in experiment.
We show in the Appendix A that a metal-insulator transition
does occur in Sr$_2$VO$_4$ with an increasing Hubbard $U_{\textrm{V}}$
and a fixed $p$-$d$ separation (via $U'_{\textrm{V}}$). However, the
critical $U_{\textrm{V}}$ is larger than typical values employed
previously in literature for the vanadates~\cite{Nekrasov-PRB-2005,
  Aichhorn-PRB-2009}.  It is possible that the experimentally observed
narrow-gap insulating behavior (experimentally observed to persist
above the N\'{e}el temperature~\cite{Matsuno-PRL-2005,
  Teyssier-PRB-2011}) arises from long-range magnetic correlations and
spatial correlations that are not
captured in our single-site paramagnetic DMFT calculation.  These
correlations relate to low energy scale physics~\cite{Gull-PRB-2010}
and are not expected to affect the charge transfer energetics of
interest here.

\subsection{Bulk manganites}

Next we discuss the bulk manganites: SrMnO$_3$ and Sr$_2$MnO$_4$.  For
ease of comparison with the superlattice results to be shown in the
next section, we study here the cubic phase of SrMnO$_3$
(isostructural to SrVO$_3$) with the lattice constant of $a$ =
3.801~\AA~(though other structures of SrMnO$_3$ also
co-exist)~\cite{Sondena-PRB-2007}. Sr$_2$MnO$_4$ forms the $n=1$
Ruddlesden-Popper structure with in-plane and out-of-plane lattice
constants $a$ = 3.802~\AA~and $c$ =
12.519~\AA~\cite{Tezuka-JSSC-1999}. Consistent with the experimental
estimation of Hubbard $U$ from photoemission
data~\cite{Park-PRL-1996}, we use a Hubbard $U_{\textrm{Mn}}$ = 5 eV
on both materials to include correlation effects on Mn $d$ orbitals.

Fig.~\ref{fig:bulk} shows the orbitally-resolved spectral function
$A(\omega)$ of bulk SrMnO$_3$ (Fig.~\ref{fig:bulk}\textbf{C}) and
Sr$_2$MnO$_4$ (Fig.~\ref{fig:bulk}\textbf{D})~\cite{manganites}.  The
threshold of O $p$ states is around 1 eV below the Fermi level. We
find that $U'_{\textrm{Mn}}$ = 4.5 eV provides a good agreement
between the calculated O $p$ states and experimental photoemission
data.  We observe that for these parameters the occupied Mn $t_{2g}$
states are visible as a peak slightly above the leading edge of the
oxygen band. We will show in Appendix B that modest changes of
parameters will move this peak slightly down in energy so that it
merges with the leading edge of the oxygen $p$ states. The experimental
situation is not completely clear. Published x-ray photoelectron
spectroscopy work ~\cite{Saitoh-PRB-1995, Zampieri-PRB-1998} indicates
a resolvable $t_{2g}$ peak at or slightly above the leading edge of
the oxygen bands; other studies including recent photoemission
measurements \cite{Saitoh-JJAP-1993, cornell} do not find a separately
resolved $t_{2g}$ peak. The issue is not important for the results of
this paper but further investigation of the location of the $t_{2g}$
states would be of interest as a way to refine our knowledge of the
electronic structure of the manganites.  With this value of
$U'_{\textrm{Mn}}$, the theory produces a small energy gap around 0.5
eV in both SrMnO$_3$ and Sr$_2$MnO$_4$.  However, the gap value is
$U_{\textrm{Mn}}$-dependent.  We show in Appendix B that with the
$p$-$d$ separation fixed, via the adjustment of $U'_{\textrm{Mn}}$, a
larger $U_{\textrm{Mn}}$ increases the Mott gap by further separating
the Mn lower and upper Hubbard bands. However, for the value of
$U_{\textrm{Mn}}$ (around 5 eV) that is extracted from photoemission
experiments ~\cite{Park-PRL-1996}, the size of the Mott gap of
Sr$_2$MnO$_4$ is substantially underestimated, compared to the optical
gap (around 2 eV) in experiment~\cite{Matsuno-PRL-2005}.  This
discrepancy may arise because this calculation does not take into
account spatial correlation~\cite{Go-2013}. However, the Mott gap is
separated by Mn $t_{2g}$ and $e_g$ states, while the energy difference
between O $p$ states and Mn $e_g$ states (i.e. $p$-$d$ separation) is
fixed by the experimental photoemission data (via
$U'_{\textrm{Mn}}$). We will show in the next section as well as in
the Appendix B that it is the $p$-$d$ separation that controls the
charge transfer and therefore the underestimation of the Mott gap does
not significantly affect our main results.

\section{Vanadate-manganite superlattices}

In this section we discuss vanadate-manganite superlattices. There are
two types: we refer to SrVO$_3$/SrMnO$_3$ superlattice as 113-type and
refer to Sr$_2$VO$_4$/Sr$_2$MnO$_4$ superlattice as 214-type. The two
types of superlattices have similarities and differences. In both
types, the charge transfer from V sites to Mn sites occurs, in which
electron dopes the Mn $e_g$ states and drains the V $t_{2g}$ states at
the Fermi level. However, in the 214 type, the VO$_6$ and MnO$_6$
octahedra are decoupled and the charge transfer arises mainly from
the electronegativity difference between V and Mn elements. In the 113
type, in addition to the electronegativity difference between V and
Mn, the movement of the shared apical oxygen changes the hybridization
and thus also affects the charge transfer. We will show below that due
to the movement of the shared apical oxygen atom, the 113-type
superlattice generically has a more enhanced charge transfer than the
214-type superlattice.

We discuss the phenomena of charge transfer in terms of: 1) structural
properties, 2) electronic properties and 3) direct electron counting.

\subsection{Structural properties}

Table~\ref{tab:structure} shows the DFT-LDA relaxed structure of
SrVO$_3$/SrMnO$_3$ and Sr$_2$VO$_4$/Sr$_2$MnO$_4$ superlattices as
well as the bulk materials.  We see that the VO$_6$ octahedron is
smaller in the superlattice than in the bulk, while the MnO$_6$
octahedron is larger.  This is suggestive that the VO$_6$
octahedron loses electrons and the MnO$_6$ octahedron gains electrons
(i.e that internal charge transfer from V to Mn sites occurs), and this will
be quantified below.

\subsection{Electronic properties}

\begin{figure}[t!]
\includegraphics[angle=-90,width=0.98\textwidth]{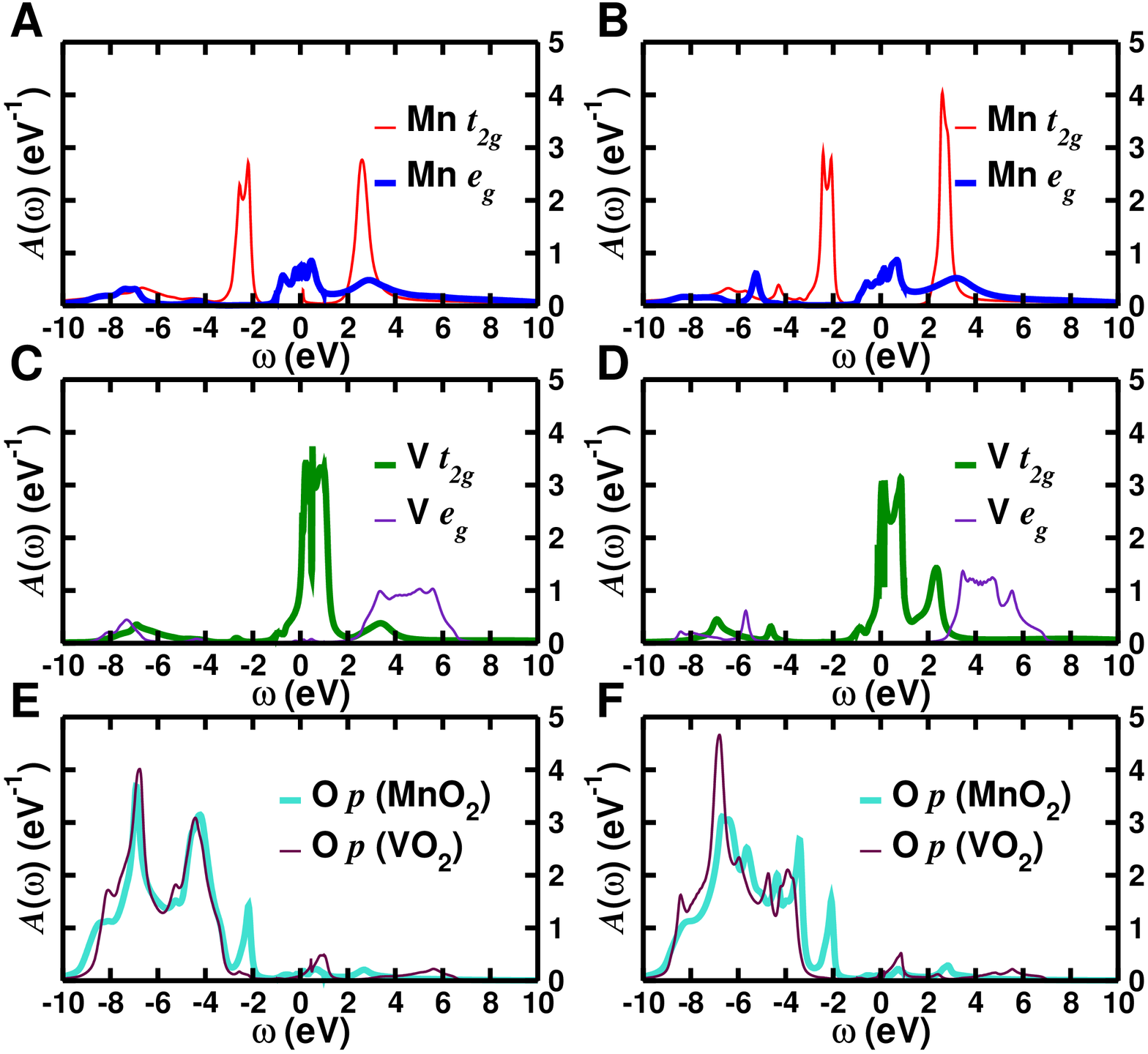}
\caption{\label{fig:SVMO} Orbitally resolved spectral function of
  vanadate-manganite superlattices, obtained from LDA+DMFT
  calculations. Left panels: SrVO$_3$/SrMnO$_3$ superlattice. Right
  panels: Sr$_2$VO$_4$/Sr$_2$MnO$_4$ superlattice.  \textbf{A}) and
  \textbf{B}): Mn $t_{2g}$ (red thin) and Mn $e_g$ (blue thick)
  states; \textbf{C}) and \textbf{D}): V $t_{2g}$ (green thick) and V
  $e_g$ (violet thin) states; \textbf{E}) and \textbf{F}): O $p$
  states of the MnO$_2$ layer (turquoise thick) and O $p$ states of
  the VO$_2$ layer (maroon thin).  $U'$ double counting is employed
  with $U_{\textrm{V}}$ = $U_{\textrm{Mn}}$ = 5 eV and
  $U'_{\textrm{V}}$ = 3.5 eV, $U'_{\textrm{Mn}}$ = 4.5 eV.  The Fermi
  level is set at zero point.}
\end{figure}

Fig.~\ref{fig:SVMO} shows the orbitally resolved spectral function
of the SrVO$_3$/SrMnO$_3$ superlattice (left panels) and the
Sr$_2$VO$_4$/Sr$_2$MnO$_4$ superlattice (right panels).  In both
superlattices, the Mn $e_g$ states emerge at the Fermi level,
while in bulk manganites, there is a small gap in the Mn $d$
states (separated by Mn $e_g$ and $t_{2g}$) in both materials. In
the VO$_2$ layer, V $t_{2g}$ states dominate at the Fermi level.
Another feature worth noting is the O $p$ states of the MnO$_2$
and of the VO$_2$ layers. Though the very first peak of O $p$
states in the MnO$_2$ layer below the Fermi level is lined up with
Mn $t_{2g}$ states due to strong covalency, the main peak almost
exactly overlaps with that of O $p$ states in the VO$_2$ layer.
This supports our hypothesis in the schematic that the main peaks
of O $p$ states of the VO$_2$ and MnO$_2$ layers are aligned in
the superlattices. We need to mention that the general features in
electronic structure of the superlattices are robust for different
double counting schemes. We show in Appendix C that the standard
FLL double counting yields a very similar electronic structure of
the superlattices. We also present LDA spectra in Appendix D for
comparison to LDA+DMFT spectra.

\begin{figure}[t]
\includegraphics[angle=-90,width=0.9\textwidth]{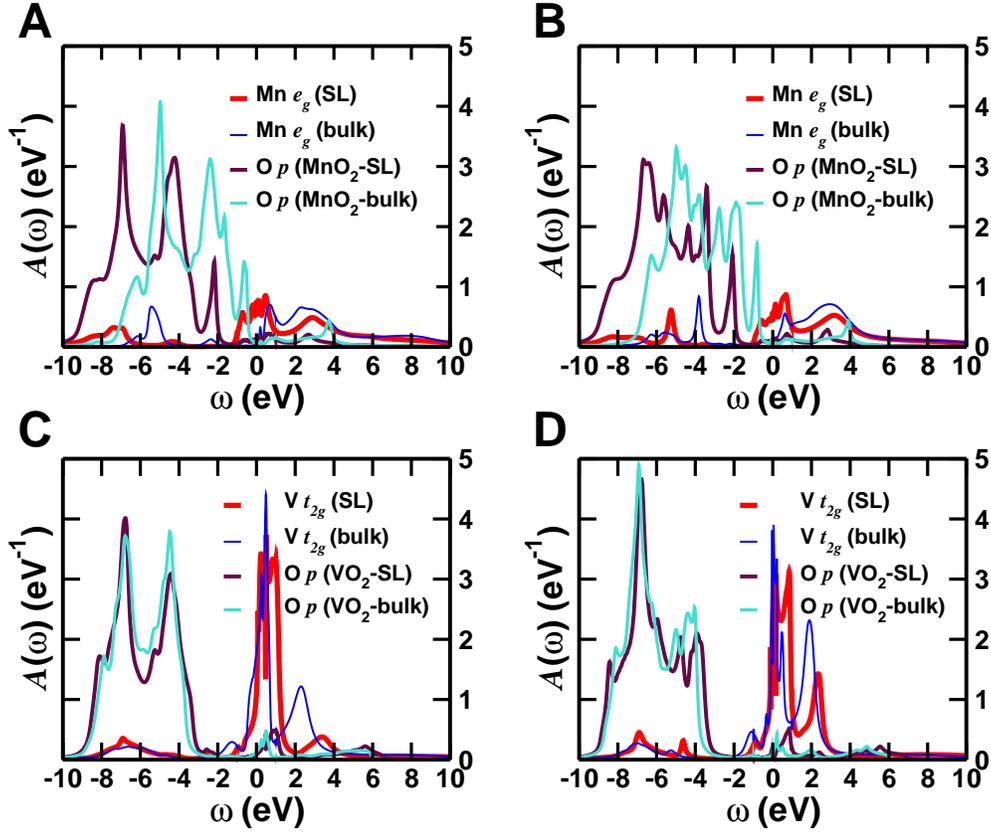}
\caption{\label{fig:compare} \textbf{A}) Comparison of Mn $e_g$ and
O $p$ states of the MnO$_2$ layer between the SrVO$_3$/SrMnO$_3$
superlattice and bulk SrMnO$_3$. \textbf{B}) Comparison of Mn $e_g$ and
O $p$ states of the MnO$_2$ layer between the Sr$_2$VO$_4$/Sr$_2$MnO$_4$
superlattice and bulk Sr$_2$MnO$_4$. \textbf{C}) Comparison of V $t_{2g}$
and O $p$ states of the VO$_2$ layer between the SrVO$_3$/SrMnO$_3$
superlattice and bulk SrVO$_3$. \textbf{D}) Comparison of V $t_{2g}$
and O $p$ states of the VO$_2$ layer between the Sr$_2$VO$_4$/Sr$_2$MnO$_4$
superlattice and bulk Sr$_2$VO$_4$. The Fermi level is set at zero energy.
``SL'' refers to the superlattices. All the spectra are obtained from
LDA+DMFT calculations.}
\end{figure}

Next, we compare the V $t_{2g}$ and Mn $e_g$ states between the
superlattices and bulk materials to show how the Fermi level shifts in
the two constituents.  Fig.~\ref{fig:compare}\textbf{A} and \textbf{B}
show the comparison of Mn $e_g$ and O $p$ states of the MnO$_2$ layer
between the superlattices and bulk manganites (\textbf{A}: 113-type
and \textbf{B}: 214-type).  The Fermi levels of bulk manganites and of
the superlattices are lined up in the same figure.

According to the schematic (Fig.~\ref{fig:band}), with respect to bulk
manganites, both the Mn $d$ and O $p$ states in the MnO$_2$ layer are
shifted towards the low energy-lying region due to the electron
doping. Fig.~\ref{fig:compare}\textbf{A} and \textbf{B} clearly
reproduce this rigid shift in i) Mn $e_g$ states from the bulk (blue
or thin dark curves) to the superlattice (red or thick light) and ii)
in O $p$ states of the MnO$_2$ layer from the bulk (turquoise or thin
light) to the superlattice (maroon or thick dark).

Similarly, Fig.~\ref{fig:compare}\textbf{C} and \textbf{D} show the
comparison of V $t_{2g}$ and O $p$ states (of the VO$_2$ layer)
between the superlattices and bulk vanadates (\textbf{C}: 113-type and
\textbf{D}: 214-type). The Fermi levels of bulk vanadates and of the
superlattices are lined up in the same figure.  According to the
schematic (Fig.~\ref{fig:band}), since electrons are drained out of V
$t_{2g}$ state, both the V $t_{2g}$ states and O $p$ states of the
VO$_2$ layer are shifted towards the high energy-lying region,
compared to their counterparts in bulk vanadates. This shift can be
seen (Fig.~\ref{fig:compare}\textbf{C} and \textbf{D}) i) in the V
$t_{2g}$ states from the bulk (blue or thin dark curves) to the
superlattice (red or thick light) and ii) in the O $p$ states of the
VO$_2$ layer from the bulk (turquoise or thin light) to the superlattice
(maroon or thick dark). However, since the peak of V $t_{2g}$ states
at the Fermi level is much higher than that of Mn $e_g$ states, the
shift in the V $t_{2g}$ states is much smaller than that in the Mn
$e_g$ states. Fig.~\ref{fig:compare} reproduces our schematic of how V
$t_{2g}$ and Mn $e_g$ states are shifted and re-arranged to reach one
common Fermi level in a vanadate-manganite superlattice. A possible
consequence is electron (hole) conductance in the MnO$_2$ (VO$_2$)
layer.

\subsection{Direct electron counting}

\begin{figure}[t!]
\includegraphics[angle=-90,width=0.9\textwidth]{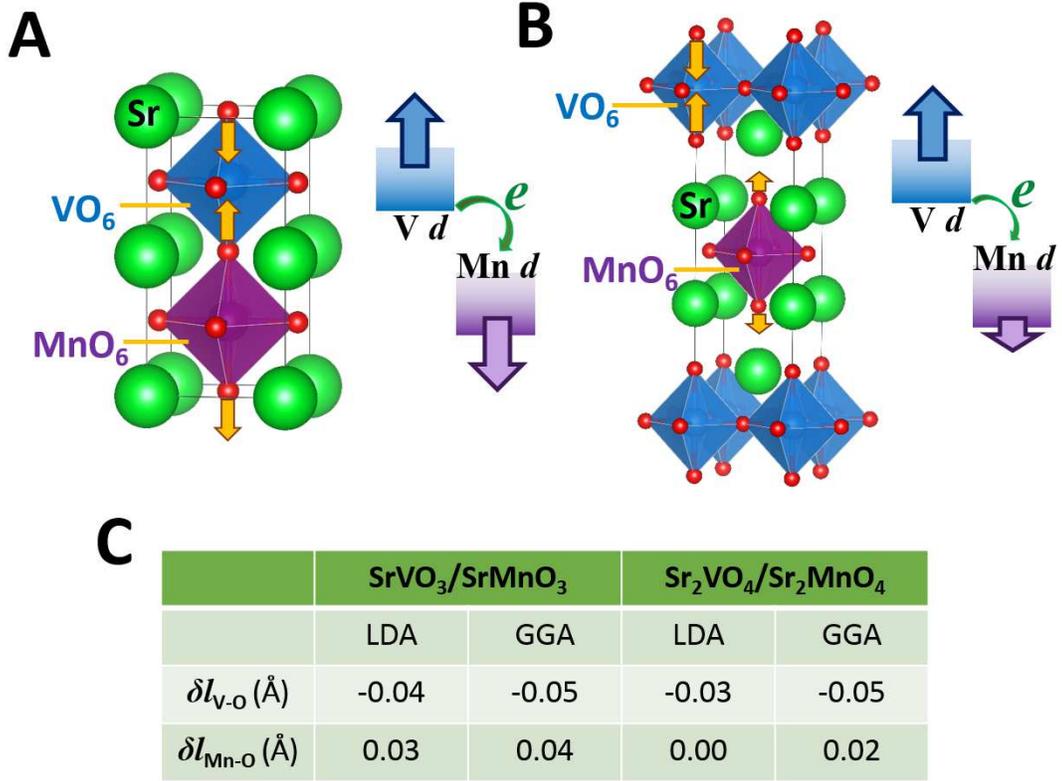}
\caption{\label{fig:hybridization} Movement of the apical oxygen,
  corresonding changes in the energy of V $d$ and Mn $d$ states and
  enhancement of the charge transfer. \textbf{A}) SrVO$_3$/SrMnO$_3$
  superlattice and \textbf{B}) Sr$_2$VO$_4$/Sr$_2$MnO$_4$
  superlattice.  The green atoms are Sr. The blue and purple cages are
  VO$_6$ and MnO$_6$ octahedra, respectively. The arrows on the oxygen
  atoms indicate the atom movement.  The arrows on the metal $d$
  states indicate the trend of energy shift. The length of the arrows
  is schematically proportional to the magnitude. \textbf{C}) Table of the
  changes of out-of-plane V-O and Mn-O bonds ($\delta l_{\textrm{V-O}}$ and
  $\delta l_{\textrm{Mn-O}}$) from bulk materials to the superlattices.}
\end{figure}

Now we calculate the occupancy on each orbital by performing the
following integral:

\begin{equation}
\label{eqn:occupancy} N_i = \int^{\infty}_{-\infty}A_i(\omega)n_F(\omega)d\omega
\end{equation}
where $A_i(\omega)$ is the spectral function for the $i$th orbital
(defined from the Wannier construction), which is defined in
Eq.~(\ref{eqn:spectral}). $n_F(\omega)$ is the fermion occupancy
factor. In order to explicitly display the charge transfer phenomenon,
we calculate the V $d$ and Mn $d$ occupancy in both bulk materials and
the superlattices. We summarize the results in Table~\ref{tab:Nd}. We
can see that $N_d$(V) decreases and $N_d$(Mn) increases from bulk to
the superlattices and an average charge transfer from V to Mn is
0.40$e$ for the 113-type superlattice and 0.25$e$ for the 214-type
superlattice.  Moreover, due to the strong covalency between
transition metal $d$ states and oxygen $p$ states, the occupancy of
oxygen $p$ states also changes between bulk materials and the
superlattices. For this reason, the change in $d$ occupancy may not be
an accurate representation of charge transfer.

We also calculate the total occupancy of VO$_2$ and MnO$_2$ layers and
find that the total charge transfer between the two layers amounts to
0.53 for the 113-type superlattice and 0.38 for the 214-type
superlattice. Unlike the 113-type superlattice in which the apical
oxygen is shared by two octahedra, the 214-type superlattice has a
unique property that each octahedron is decoupled between
layers. Therefore in the superlattice, we can count the charge
transfer from the VO$_6$ octahedron to the MnO$_6$ octahedron. Note
that since we only take into account the $p$-$d$ band manifold, the V
and Mn octahedra include all the Wannier states and therefore in bulk
Sr$_2$VO$_4$, the number of electrons per VO$_4$ unit is exactly 25$e$
and in bulk Sr$_2$MnO$_4$, the number of electrons per MnO$_4$ unit is
exactly 27$e$. We find that relative to the bulk materials, the V
octahedron of the 214-type superlattice loses 0.48$e$ and Mn octahedron
of the 214-type superlattice gains exactly 0.48$e$.  Comparison of
this 0.48$e$ charge transfer to the 0.25$e$ found by only considering
$d$ orbitals further confirms that not only the transition metal $d$
states but also oxygen $p$ states participate in the charge
transfer. From Table~\ref{tab:Nd}, we can see that the internal charge
transfer is stronger in the 113-type superlattice, compared to the
214-type. We show below that the difference arises because in the
113-type superlattice, the apical oxygen is shared by the VO$_6$ and
MnO$_6$ octahedra, whereas the octahedra are decoupled in the
214-type.

We see from Table~\ref{tab:structure}) that due to the internal charge
transfer, the VO$_6$ octahedron loses electrons and shrinks; on the
other hand, the MnO$_6$ octahedron gains electrons and
expands. Therefore the shared apical oxygen atom moves away from Mn
sites and towards V sites (see
Fig.~\ref{fig:hybridization}\textbf{A}). A direct consequence is that
the out-of-plane Mn-O hopping decreases and the out-of-plane V-O
hopping increases. Since the V $d$ and Mn $d$ states are anti-bonding
in nature, the changes in the metal-ligand hopping push the V $d$
states higher in energy and lower the energy of Mn $d$ states and thus
enhance the internal charge transfer.  In the 214-type superlattice,
we have a different situation because the two oxygen octahedra have
their own apical oxygen atoms, whose movements are decoupled. From the
Table~\ref{tab:structure} and Fig.~\ref{fig:hybridization}\textbf{C},
the VO$_6$ shrinks and the apical oxygen
atom of VO$_6$ moves towards the V atom, just like the 113-type
superlattice. However, the MnO$_6$ expands but the moveoment of
apical oxygen is much smaller (the in-plane Mn-O bond does increase,
so does the overall volume of MnO$_6$). Therefore, the energy of V $d$ states is
increased due to the enhanced out-of-plane V-O hopping, but the energy
of Mn $d$ states does not decrease much because the movement of apical oxygen
atom is reduced (Fig.~\ref{fig:hybridization}\textbf{B}). As a
result, the charge transfer between V and Mn sites is weaker in the
214-type superlattice, compared to the 113-type superlattice.

Our discussions in this paper have focussed mainly on the
(SrVO$_3$)$_1$/(SrMnO$_3$)$_1$ superlattice. Though $m=1$
superlattices (in the notation of (SrVO$_3$)$_m$/(SrMnO$_3$)$_m$)
are easy for theoretical studies, experimentally it is more
practical to grow $m=2$ or larger $m$ superlattices. We show in
the Appendix E that comparing (SrVO$_3$)$_1$/(SrMnO$_3$)$_1$ and
(SrVO$_3$)$_2$/(SrMnO$_3$)$_2$ superlattices, the charge transfer
is very similar. However, for a large $m$, we will have
inquivalent V sites and eventually the charge transfer will be
confined to the interfacial region. Investigating the length
scales associated with charge transfer is an important open
question.

\begin{table}[t]
\caption{\label{tab:Nd} The occupancy of V $d$ and Mn $d$ states, as
  well as VO$_2$ and MnO$_2$ layers in vanadates, manganites and the
  superlattices. All the occupancies are calculated from Wannier basis
  using the DFT-LDA or DFT-GGA relaxed structures. $\overline{\Delta
    N_d}$ ($\overline{\Delta N}$)~\cite{delta} is the average charge
  transfer between V $d$ and Mn $d$ states (VO$_2$ and MnO$_2$ layers,
  or VO$_4$ and MnO$_4$ octahedra), using the DFT-LDA relaxed
  structures.}
\begin{center}
\begin{tabular}{c|c|c|c|c}
\hline
\hline
SrVO$_3$ & SrMnO$_3$ & \multicolumn{2}{c|}{SrVO$_3$/SrMnO$_3$} &  \\
\hline
LDA/GGA & LDA/GGA & LDA/GGA & LDA/GGA & LDA/GGA \\
\hline
$N_d$(V) & $N_d$(Mn) & $N_d$(V) & $N_d$(Mn) & $\overline{\Delta N_d}$\\
2.09/2.01   & 4.08/4.05   &  1.73/1.59 &  4.51/4.57 & 0.40/0.47 \\
\hline
$N$(VO$_2$) & $N$(MnO$_2$) & $N$(VO$_2$) & $N$(MnO$_2$) & $\overline{\Delta N}$ \\
13.36/13.34 & 15.36/15.35 & 12.86/12.73 & 15.92/16.03 & 0.53/0.65 \\
\hline
\hline
Sr$_2$VO$_4$ & Sr$_2$MnO$_4$ & \multicolumn{2}{c|}{Sr$_2$VO$_4$/Sr$_2$MnO$_4$} &  \\
\hline
LDA/GGA  &  LDA/GGA &  LDA/GGA & LDA/GGA & LDA/GGA \\
\hline
$N_d$(V) & $N_d$(Mn) & $N_d$(V) & $N_d$(Mn) & $\overline{\Delta N_d}$\\
2.07/2.00 & 4.12/4.11 & 1.86/1.71 & 4.41/4.44 & 0.25/0.31 \\
\hline
$N$(VO$_2$) & $N$(MnO$_2$) & $N$(VO$_2$) & $N$(MnO$_2$) & $\overline{\Delta N}$ \\
13.35/13.34 & 15.42/15.44 & 13.01/12.90 &  15.83/15.91 & 0.38/0.46 \\
\hline
$N$(VO$_4$) & $N$(MnO$_4$) & $N$(VO$_4$) & $N$(MnO$_4$) & $\overline{\Delta N}$ \\
25.00/25.00 & 27.00/27.00 & 24.52/24.42 & 27.48/27.58 & 0.48/0.58 \\
\hline\hline
\end{tabular}
\end{center}
\end{table}

\section{Conclusions}

We use DFT+DMFT calculations to show that due to the difference in
electronegativity, internal charge transfer could occur between
isostructural vanadates and manganites in both 113-type and
214-type superlattices. The charge transfer is enhanced by
associated lattice distortions.  The moderate electronegativity
difference between Mn and V leads to moderate charge transfer, in
contrast to the LaTiO$_3$/LaNiO$_3$ superlattice, in which a
complete charge transfer fills up the holes on the oxygen atoms in
the NiO$_2$ layer~\cite{electroneg}. The partially filled bands
imply metallic conductance that could possibly be observed in
transport, if the thin film quality is high enough that disorder
is suppressed and Anderson localization does not
occur~\cite{Lee-Ramakrishnan-RMP-1985}. Our study of a
superlattice consisting of two different species of transition
metal oxides establishes that internal charge transfer is a
powerful tool to engineer electronic structure and tailor
correlation effects in transition metal oxides~\cite{Chen-PRL-2013b, 
Kleibeuker-PRL-2014}. In particular, for
vanadate-manganite superlattices, internal charge transfer may
serve as an alternative approach to dope Mott insulators without
introducing chemical disorder. Furthermore, as previous works have
shown~\cite{Akamatsu-PRL-2014, Mulder-AdvFuncMater-2013}, in
addition to perovskite structure, Ruddlesden-Popper structures can
also be an important ingredient in the design of oxide
superlattices with tailored properties~\cite{Lee-Nature-2013}.
Finally, our examination of different materials raises the issue of
the value of the double counting coefficient $U'$, determined
here by fitting photoemission data. Understanding the variation of
$U'$ across the transition metal oxide family of materials is an
important open problem.

\begin{acknowledgments}
We are grateful to Darrell Schlom, Kyle Shen, Eric Monkman and Masaki
Uchida for stimulating discussion and sharing the unpublished data
with us. H. Chen is supported by National Science Foundation under
Grant No. DMR-1120296. A. J. Millis is supported by the Department of
Energy under Grant No. DOE-ER-046169. H. Park and C. A. Marianetti are
supported by FAME, one of six centers of STARnet, a Semiconductor
Research Corporation program sponsored by MARCO and DARPA.
Computational facilities are provided via XSEDE resources through
Grant No. TG-PHY130003.
\end{acknowledgments}

\appendix

\clearpage
\newpage

\section{Metal-insulator transition of S\lowercase{r}$_2$VO$_4$}

\begin{figure}[h!]
\includegraphics[angle=-90,width=0.9\textwidth]{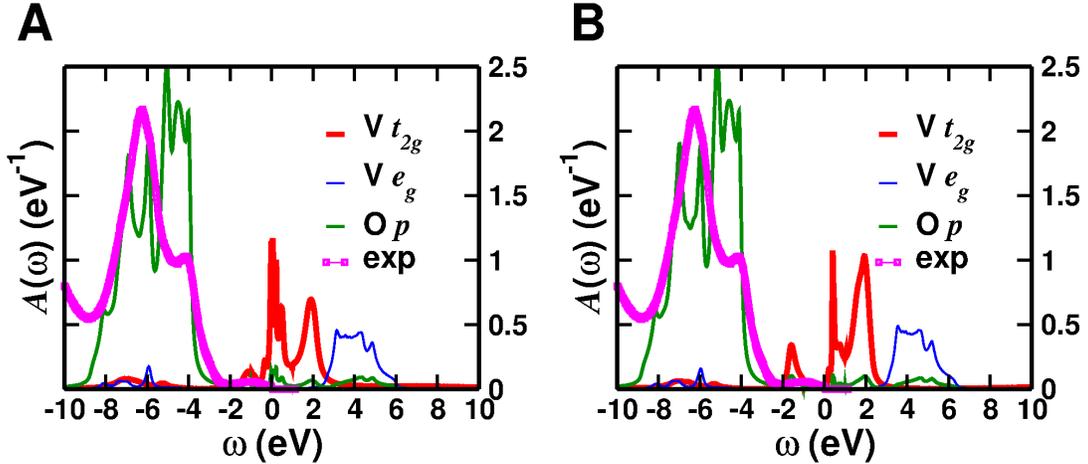}
\caption{\label{fig:appendix-A} Orbitally resolved spectral function
  of Sr$_2$VO$_4$, obtained from LDA+DMFT calculations. $U'$ double
  counting is employed with \textbf{A}) $U_{\textrm{V}}$ = 5 eV,
  $U'_{\textrm{V}}$ = 3.5 eV and \textbf{B}) with $U_{\textrm{V}}$ = 8
  eV, $U'_{\textrm{V}}$ = 6.8 eV.  The red (very thick), blue (thick)
  and green (thin) curves are V $t_{2g}$, V $e_g$ and O $p$ projected
  spectral functions, respectively.  The pink dots are experimental
  photoemission data for SrVO$_3$~\cite{cornell}.  The Fermi level is
  set at zero energy.}
\end{figure}

In this appendix, we show that within single-site DMFT and with the
$p$-$d$ separation fixed by the experimental photoemission data, there
is a metal-insulator transition in Sr$_2$VO$_4$ with an increasing
Hubbard $U_{\textrm{V}}$ ($U'_{\textrm{V}}$ is determined by the
$p$-$d$ separation for each given
$U_{\textrm{V}}$). Fig.~\ref{fig:appendix-A}\textbf{A} shows that for
$U_{\textrm{V}}$ = 5 eV and $U'_{\textrm{V}}$ = 3.5 eV, Sr$_2$VO$_4$
is metallic with mainly V $t_{2g}$ states at the Fermi level, which is
a reproduction of
Fig~\ref{fig:bulk}\textbf{B}. Fig.~\ref{fig:appendix-A}\textbf{B}
shows that with $U_{\textrm{V}}$ increased to 8 eV and
$U'_{\textrm{V}}$ to 6.8 eV which approximately fixes the $p$-$d$
separation, a metal-insulator transition occurs and Sr$_2$VO$_4$ is
rendered a Mott insulator. However, the critical $U_{\textrm{V}}$
depends on the approximation scheme we employ.  A more elaborate
cluster-DMFT calculation and/or the inclusion of long range order may
find a smaller critical $U_{\textrm{V}}$~\cite{Go-2013}.

\clearpage
\newpage

\section{Mott gap of S\lowercase{r}$_2$M\lowercase{n}O$_4$ and its effects on
S\lowercase{r}$_2$M\lowercase{n}O$_4$/S\lowercase{r}$_2$VO$_4$
superlattices}

\begin{figure}[h!]
\includegraphics[angle=-90,width=0.9\textwidth]{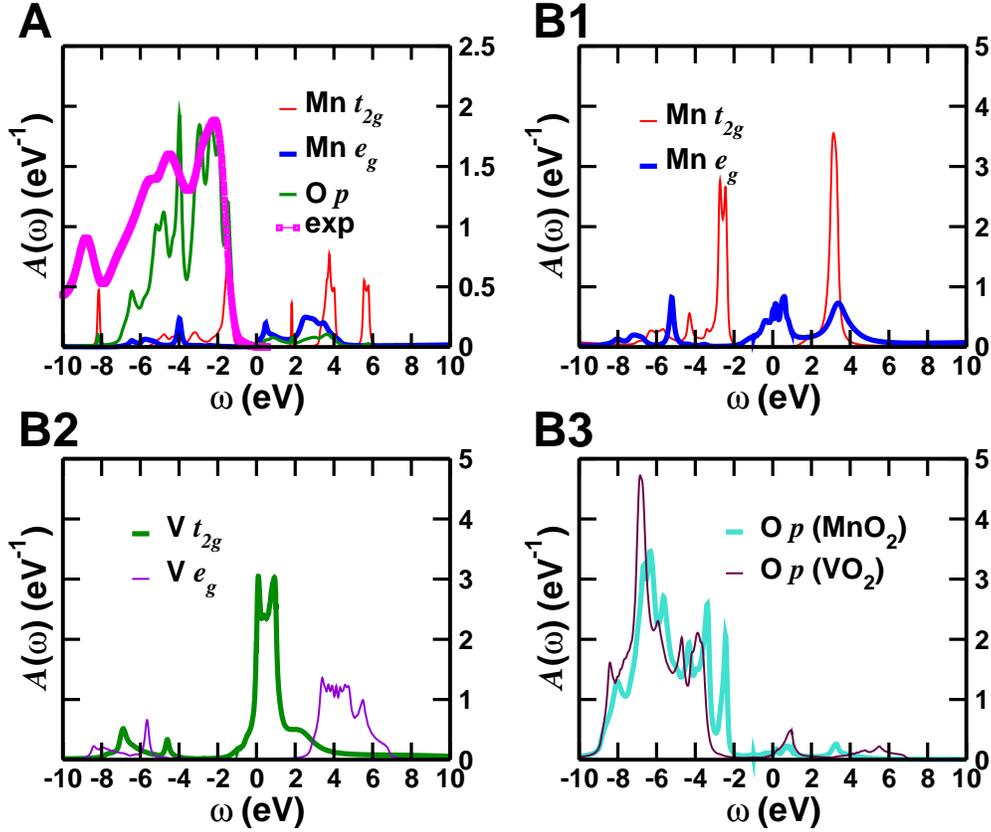}
\caption{\label{fig:appendix-B} \textbf{A}) Orbitally resolved
  spectral function of Sr$_2$MnO$_4$, obtained from LDA+DMFT
  calculations.  $U'$ double counting is employed with
  $U_{\textrm{Mn}}$ = 8 eV, $U'_{\textrm{Mn}}$ = 7.5 eV. The red
  (thin), blue (very thick) and green curves (thick) are Mn $t_{2g}$,
  Mn $e_g$ and O $p$ projected spectral functions, respectively.  The
  pink dots are experimental photoemission data of
  SrMnO$_3$~\cite{cornell}.  The Fermi level is set at zero energy.
  \textbf{B}) Orbitally resolved spectral function of
  Sr$_2$VO$_4$/Sr$_2$MnO$_4$ superlattices, obtaine from LDA+DMFT
  calculations.  $U'$ double counting is employed with
  $U_{\textrm{Mn}}$ = 8 eV, $U'_{\textrm{Mn}}$ = 7.5 eV and
  $U_{\textrm{V}}$ = 5 eV, $U'_{\textrm{V}}$ = 3.5 eV. \textbf{B1}):
  Mn $t_{2g}$ (red thin) and Mn $e_g$ (blue thick) states;
  \textbf{B2}): V $t_{2g}$ (green thick) and V $e_g$ (violet thin)
  states; \textbf{B3}): O $p$ states of the MnO$_2$ layer (turquoise
  thick) and O $p$ states of the VO$_2$ layer (maroon thin).}
\end{figure}

In this appendix, we show how the Hubbard $U_{\textrm{Mn}}$ changes
the Mott gap of Sr$_2$MnO$_4$ with the $p$-$d$ separation
approximately fixed.  Fig.~\ref{fig:appendix-B}\textbf{A} shows the
orbitally resolved spectral function of Sr$_2$MnO$_4$ with
$U_{\textrm{Mn}} =$ 8 eV and $U'_{\textrm{Mn}} =$ 7.5 eV.
Note that since Sr$_2$MnO$_4$ is a Mott insulator, the
Fermi level in the calculation is shifted at the conduction band edge,
i.e. the edge of Mn $e_g$ states. In Fig.~\ref{fig:bulk}\textbf{D} of
the main text, the Mott gap of Sr$_2$MnO$_4$ is around 0.5 eV with
$U_{\textrm{Mn}}$ = 5 eV and $U'_{\textrm{Mn}}$ = 4.5 eV. If we
increase $U_{\textrm{Mn}}$ to 8 eV and $U'_{\textrm{Mn}}$ to 7.5 eV,
the Mott gap is correspondingly increased to around 1 eV with the
$p$-$d$ separation fixed by the photoemission data~\cite{cornell}. The
Mn $t_{2g}$ peak and the main peak of O $p$ states now merge
together. However, even with $U_{\textrm{Mn}}$ = 8 eV, the Mott gap is
still smaller than the optical gap (around 2 eV) from
experiment~\cite{Matsuno-PRL-2005}. The difference could be due to
spatial correlations not included in our single-site DMFT
approximation~\cite{Go-2013}.

Using the parameters $U_{\textrm{Mn}}$ = 8 eV and $U'_{\textrm{Mn}}$ = 7.5 eV,
we redo the calculations on Sr$_2$VO$_4$/Sr$_2$MnO$_4$ superlattices (with
$U_{\textrm{V}}$ = 5 eV and $U'_{\textrm{V}}$ = 3.5 eV) to test the effects
of Mott gap size on charge transfer. As Fig.~\ref{fig:appendix-B}\textbf{B}
shows, the key features in electronic structure remain the same as
Fig.~\ref{fig:SVMO} in the main text: i) Mn $e_g$
and V $t_{2g}$ states emerge at the Fermi level and ii) the main peaks of
O $p$ states associated with the MnO$_2$ and VO$_2$ layers are approximately
aligned. This shows that it is the $p$-$d$ separation that controls the
charge transfer across the interface while the size of Mott gap plays
a secondary role.

\clearpage
\newpage

\section{Electronic structure calculated using the fully localized limit double
counting}

\begin{figure}[h!]
\includegraphics[angle=-90,width=0.98\textwidth]{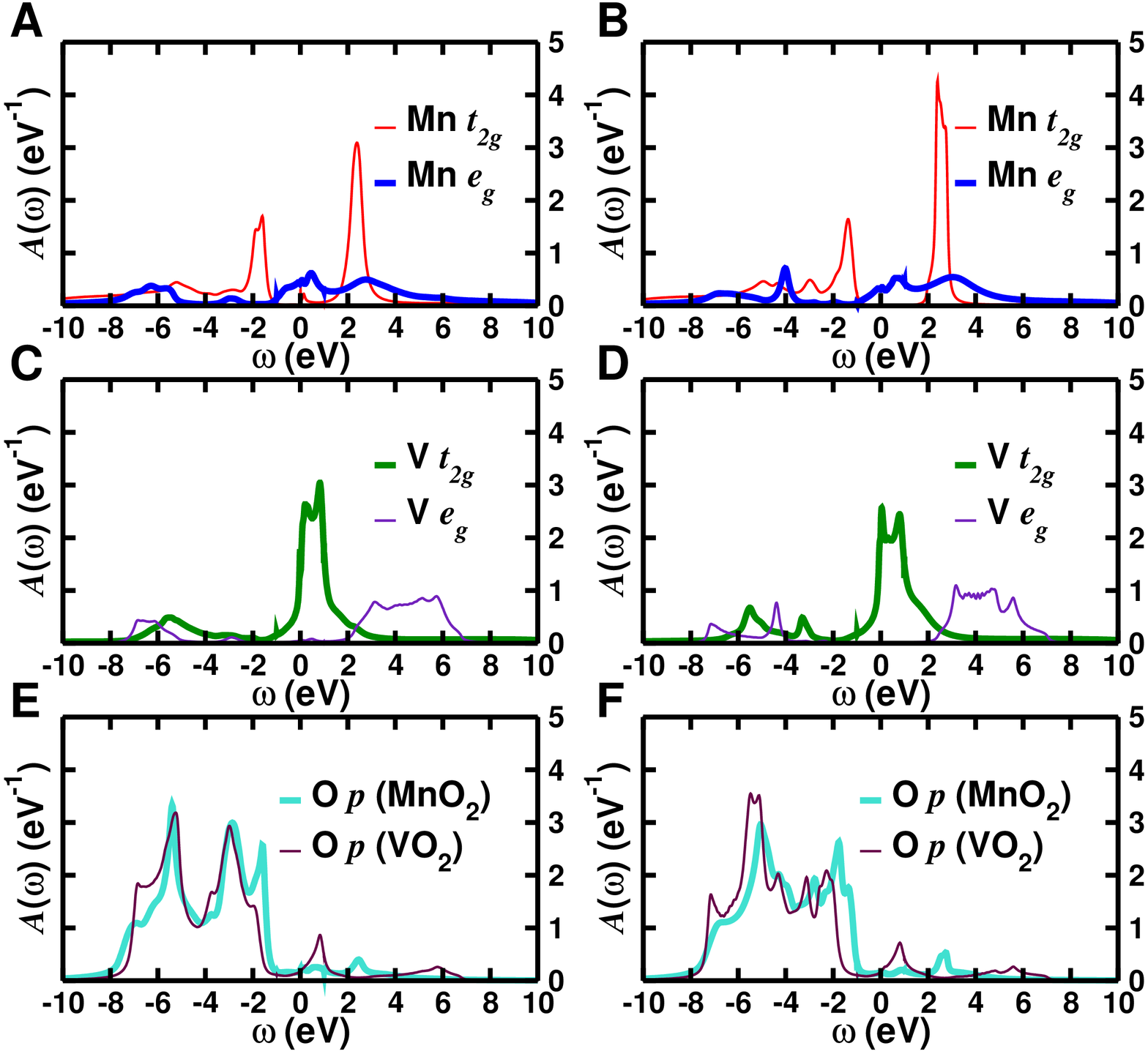}
\caption{\label{fig:FLL} Orbitally resolved spectral function of
  vanadate-manganite superlattices, obtained from LDA+DMFT
  calculations. Left panels: SrVO$_3$/SrMnO$_3$ superlattice. Right
  panels: Sr$_2$VO$_4$/Sr$_2$MnO$_4$ superlattice.  \textbf{A}) and
  \textbf{B}): Mn $t_{2g}$ (red thin) and Mn $e_g$ (blue thick)
  states; \textbf{C}) and \textbf{D}): V $t_{2g}$ (green thick) and V
  $e_g$ (violet thin) states; \textbf{E}) and \textbf{F}): O $p$
  states of the MnO$_2$ layer (turquoise thick) and O $p$ states of
  the VO$_2$ layer (maroon thin).  Fully localized limit double
  counting is employed with $U_{\textrm{V}}$ = $U_{\textrm{Mn}}$ = 5
  eV. The Fermi level is set at zero energy.}
\end{figure}

In this appendix, we show the electronic structure of both
SrVO$_3$/SrMnO$_3$ and Sr$_2$VO$_4$/Sr$_2$MnO$_4$ superlattices,
calculated using the standard fully localized limit (FLL) double
counting.  The orbitally resolved spectral function is shown in
Fig.~\ref{fig:FLL}, which is compared to Fig.~\ref{fig:SVMO} in the
main text. We employ $U_{\textrm{V}}$ = $U_{\textrm{Mn}}$ = 5 eV. We
find the FLL double counting does not change the key features of
electronic structure, such as the emergence of Mn $e_g$ and V $t_{2g}$
states at the Fermi level.

\clearpage
\newpage

\section{Electronic structure calculated using the local density approximation}

\begin{figure}[h!]
\includegraphics[angle=-90,width=0.9\textwidth]{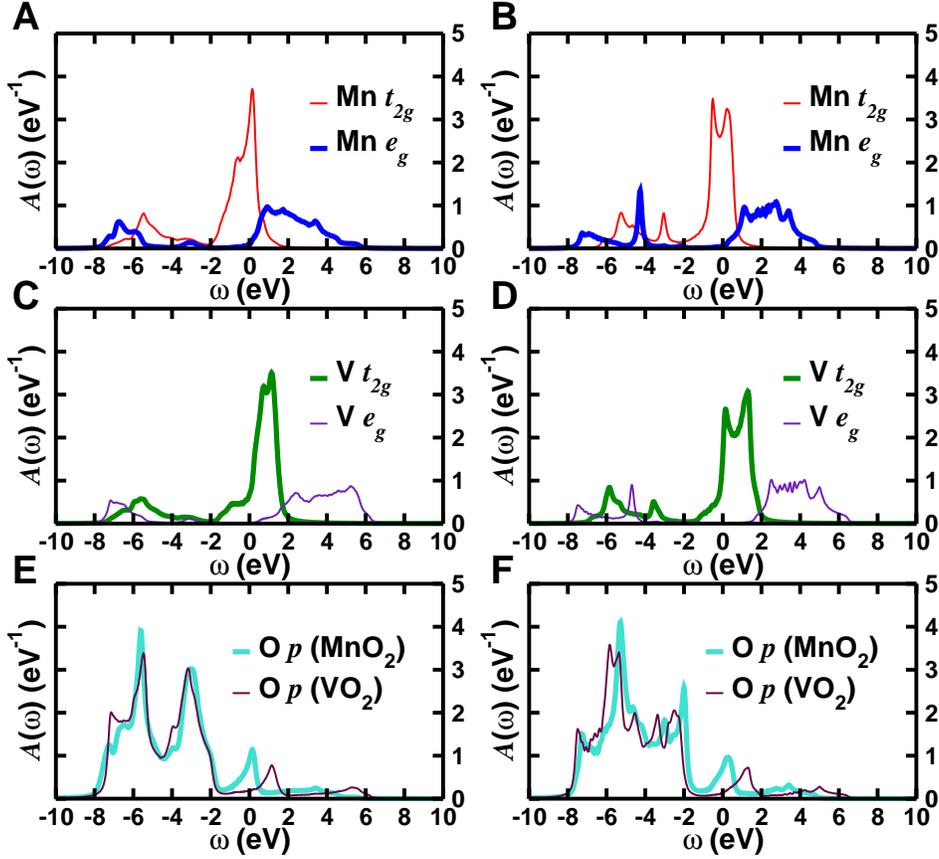}
\caption{\label{fig:LDA} Orbitally resolved spectral function of
  vanadate-manganite superlattices, obtained from LDA
  calculations. Left panels: SrVO$_3$/SrMnO$_3$ superlattice. Right
  panels: Sr$_2$VO$_4$/Sr$_2$MnO$_4$ superlattice.  \textbf{A}) and
  \textbf{B}): Mn $t_{2g}$ (red thin) and Mn $e_g$ (blue thick)
  states; \textbf{C}) and \textbf{D}): V $t_{2g}$ (green thick) and V
  $e_g$ (violet thin) states; \textbf{E}) and \textbf{F}): O $p$
  states of the MnO$_2$ layer (turquoise thick) and O $p$ states of
  the VO$_2$ layer (maroon thin).}
\end{figure}

In this appendix, we show the electronic structure of both
SrVO$_3$/SrMnO$_3$ and Sr$_2$VO$_4$/Sr$_2$MnO$_4$ superlattices,
calculated using the local density approximation alone.
The orbitally resolved spectral function is shown in
Fig.~\ref{fig:LDA}, which is compared to Fig.~\ref{fig:SVMO} in the
main text. We illustrate that without including strong correlation effect via
Hubbard $U$, Mn $t_{2g}$ states lie around the Fermi level and
do not split into lower and upper Hubbard bands.

\clearpage
\newpage

\section{(S\lowercase{r}VO$_3$)$_1$/(S\lowercase{r}M\lowercase{n}O$_3$)$_1$ versus (S\lowercase{r}VO$_3$)$_2$/(S\lowercase{r}M\lowercase{n}O$_3$)$_2$
superlattices}

In this appendix, we compare the (SrVO$_3$)$_1$/(SrMnO$_3$)$_1$
superlattice to the (SrVO$_3$)$_2$/(SrMnO$_3$)$_2$
superlattice. We focus on the $d$ occupancy and the charge transfer
from V to Mn sites. Table~\ref{tab:Nd-a} shows that $N_d$ of V sites and
Mn sites are very similar between (SrVO$_3$)$_1$/(SrMnO$_3$)$_1$ and
(SrVO$_3$)$_2$/(SrMnO$_3$)$_2$ superlattices.

\begin{table}[h!]
\caption{\label{tab:Nd-a}The occupancy of V $d$ and Mn $d$ states, as well as
VO$_2$ and MnO$_2$ layers in vanadates, manganites and the superlattices. All
the occupancies without the parentheses are calculated from Wannier basis
using the DFT-LDA relaxed structures. The occupancies in the parentheses
are calculated from Wannier basis using the experimental structures.}
\begin{center}
\begin{tabular}{c|c|c|c|c|c}
\hline
\hline
SrVO$_3$ & SrMnO$_3$ & \multicolumn{2}{c|}{(SrVO$_3$)$_1$/(SrMnO$_3$)$_1$} &  \multicolumn{2}{c}{(SrVO$_3$)$_2$/(SrMnO$_3$)$_2$}  \\
\hline
$N_d$(V) & $N_d$(Mn) & $N_d$(V) & $N_d$(Mn) &  $N_d$(V) & $N_d$(Mn) \\
2.09 (2.03) & 4.08 (4.06) &  1.73 &  4.51  &  1.69  &  4.54   \\
\hline
$N$(VO$_2$) & $N$(MnO$_2$) & $N$(VO$_2$) & $N$(MnO$_2$) &  $N$(VO$_2$) & $N$(MnO$_2$)  \\
13.36 (13.34)  &  15.36 (15.35) & 12.86 & 15.92  &  12.82 & 15.95 \\
\hline\hline
\end{tabular}
\end{center}
\end{table}

\newpage
\clearpage


\begin{thebibliography}{79}
\bibitem[{man()}]{manganites}
\bibinfo{note}{Due to the insulating nature, the theoretical Fermi level of
  both SrMnO$_3$ and Sr$_2$MnO$_4$ is shifted to the Mn conduction band edge,
  and is aligned with the experimental Fermi level when the spectral function
  and the photoemission data are compared.}
\bibitem[{del()}]{delta}
\bibinfo{note}{The explicit definition of $\overline{\Delta N_d}$ is
  $\overline{\Delta N_d} = \frac{1}{2}\Big(|N_d(\textrm{V}_{\textrm{bulk}}) -
  N_d(\textrm{V}_{\textrm{superlattice}})| + |N_d(\textrm{Mn}_{\textrm{bulk}})
  - N_d(\textrm{Mn}_{\textrm{superlattice}})|\Big)$. $\overline{\Delta N}$ uses
  a similar definition.}
\bibitem[{ele()}]{electroneg}
\bibinfo{note}{The electronegativity $\chi$ of V in SrVO$_3$ and Mn in
  SrMnO$_3$ can be estimated by employing the Wannier functions to calculate
  the $p$-$d$ separation. Among the three O $p$ states, using the one with the
  highest onsite energy, and the relevant transition metal $d$ states, we find
  that in SrVO$_3$,$\chi$(V-$t_{2g}$) = -3.13 eV and in SrMnO$_3$
  $\chi$(Mn-$e_g$) = -2.99 eV. Therefore $\chi$(Mn-$e_g$) - $\chi$(V-$t_{2g}$)
  = 0.14 eV $>$ 0, leading to a moderate charge transfer from V to Mn. Applying
  the same methods to LaNiO$_3$ and LaTiO$_3$, we find that $\chi$(Ni-$e_g$) -
  $\chi$(Ti-$t_{2g}$) = 2.93 eV, leading a complete charge transfer from Ti to
  Ni.}
\end{thebibliography}

\end{document}